\documentclass[conference]{IEEEtran}
\IEEEoverridecommandlockouts
\usepackage[left=1.62cm,right=1.62cm,top=1.9cm]{geometry}
\usepackage{algorithm}
\usepackage{url}
\usepackage{float}
\usepackage{wrapfig}
\usepackage{cite}
\usepackage{amsmath,amssymb,amsfonts}
\usepackage{algorithmic}
\usepackage[algo2e]{algorithm2e} 
\usepackage{graphicx}
\usepackage{textcomp}
\usepackage{xcolor}

\def\BibTeX{{\rm B\kern-.05em{\sc i\kern-.025em b}\kern-.08em
    T\kern-.1667em\lower.7ex\hbox{E}\kern-.125emX}}
\begin{document}

\title{A Time Series Forecasting Approach to Minimize Cold Start Time in Cloud-Serverless Platform}

\author{\IEEEauthorblockN{Akash Puliyadi Jegannathan\IEEEauthorrefmark{4}, Rounak Saha\IEEEauthorrefmark{4}\thanks{\IEEEauthorrefmark{4} Equal contribution} and Sourav Kanti Addya}
	\IEEEauthorblockA{Department of Computer Science and Engineering\\
		National Institute of Technology Karnataka, Surathkal, India\\
		Email: \{pjakash10,rounak.rs69\}@gmail.com, souravkaddya@nitk.edu.in}}

\maketitle

\begin{abstract}
Serverless computing is a buzzword that is being used commonly in the world of technology and among developers and businesses. Using the Function-as-a-Service (FaaS) model of serverless, one can easily deploy their applications to the cloud and go live in a matter of days, it facilitates the developers to focus on their core business logic and the backend process such as managing the infrastructure, scaling of the application, updation of software and other dependencies is handled by the Cloud Service Provider. One of the features of serverless computing is ability to scale the containers to zero, which results in a problem called cold start. The challenging part is to reduce the cold start latency without the consumption of extra resources. In this paper, we use SARIMA (Seasonal Auto Regressive Integrated Moving Average), one of the classical time series forecasting models to predict the time at which the incoming request comes, and accordingly increase or decrease the amount of required containers to minimize the resource wastage, thus reducing the function launching time. Finally, we implement PBA (Prediction Based Autoscaler) and compare it with the default HPA (Horizontal Pod Autoscaler), which comes inbuilt with kubernetes. The results showed that PBA performs fairly better than the default HPA, while reducing the wastage of resources.
\end{abstract}

\section{Introduction}
Serverless Computing is a state-of-the-art Cloud Computing paradigm that engrosses tremendous attention from industry and academia due to its architectonics. Serverless Computing does not mean that there are no servers; it merely means that the servers are there, but the user does not have the hassle to manage those\cite{baldini2017serverless}. In Serverless computing, applications or microservices are deployed as functions that start working when any event has triggered it. During the function life-cycle, all the operational overheads are managed by the cloud service provider (CSP). Deploying multi-tier application \cite{addya2021comcloud} become easier for CSP using the concept of inter dependent function in serverless computing.  

Serverless Computing comes with many advantages. Significant of them are no back-end operational overheads means the CSP does all back-end-related stuff such as server management, auto-scaling, load-balancing and others so that the users focus only on developing the business logic and granular on-demand pricing in order of milliseconds, which means the execution cost of any function charged on the millisecond scale \cite{kim2020automated}. Due to this flexibility in service-deployment, there is a prolific shift of users towards serverless computing seen in the recent times, especially for applications such as deep learning \cite{ishakian2018serving}, blockchain \cite{benedict2020serverless}, Internet of Things \cite{wang2020supporting}, big data analytics \cite{rahman2019serverless}, and others\cite{baresi2019towards}, \cite{wang2019distributed}. Therefore, major cloud service providers such as AWS, Microsoft Azure, Google Cloud, and IBM Cloud came up with their serverless platform named \texttt{AWS Lambda}\footnote{https://aws.amazon.com/lambda/}, \texttt{Azure Function}\footnote{https://azure.microsoft.com/en-in/services/functions/}, \texttt{GCP Function}\footnote{https://cloud.google.com/functions}, and \texttt{IBM Function}\footnote{https://cloud.ibm.com/functions} to gratify this large faction, respectively.

However, despite having much prosperity, Serverless Computing has several drawbacks that deteriorate its performance at an extensive level; the Cold-start problem is one of the major of them. The definition of the Cold-start problem can be derived from the function life-cycle. In serverless, the function life cycle starts when an event triggers it; after that, containers are created according to the function's requirements, then the environment and network are set up according to the functions image file finally, the source code files of the function is loaded, and the function starts execution. The phase that starts after triggering the function until before execution is called containerization \cite{lin2020modeling} and the time required for it is majorly called cold-start time. The problem of minimizing this cold-start time is called as the cold-start problem. A recent study done by \cite{vahidinia2020cold} showcases that in severe circumstances, the cold-start time can go up to 59 times higher than the actual time of execution. These circumstances occurred for many reasons; the unavailability of computing resources is one of the major of them \cite{manner2018cold}. In serverless, auto-scaling of resources is one of the significant features that when the incoming web traffic load is getting low, resources are scaled down according to the need. However, this is the leading cause of the unavailability of computing resources. Whenever resource requirements increase rapidly w.r.t actual scaling of resources, lesser resources need to handle heavy workloads hence containerization process gets prolonged; hence cold-start problem gets severe.Various serverless platforms offer to capture the concealed computing resource information, and some open-source serverless frameworks use this to take advantage of anticipation resource scaling. The most used and optimized resource auto-scaling is Horizontal Pod AutoScaler (HPA) which is provided by Kubernetes \cite{Kubernetes} primitive serverless framework Kubeless. Kubeless uses underlying resource information and function configuration knowledge for resource scaling decision-making.

According to the earlier observations, the proper accessibility of resources could efficiently minimize the cold-start problem. In this work, we propose a Time Series based learning model using Seasonal Auto-Regressive Integrated Moving Average(SARIMA). This model analyses previous functions invoking data patterns. Based on that, it predicts the future incoming function invocation request and scales up/down resources correspondingly. This model also utilizes resource usage efficiently. We compare the proposed model with HPA provided by kubernetes, which is the best in class auto-scaler available. This helps in analyzing and examining the performance of both configurations.

The \textbf{key contributions} of our work are as follows:
\begin{itemize}
    \item A learning-based time series forecasting model named \textit{Prediction Based Autoscaler} (PBA) that predicts the number of future function invocations and allocates computing resources accordingly, making it possible to spin up the containers before high function invocation traffic. This also helps to reduce the number of resources that would have gone waste if there were no scaling requirements.
    \item Comparative analysis of our proposed model against HPA comes inbuilt with Kubernetes for the synthetic function workload pattern.
\end{itemize}

The paper is composed of various sections. Section II contains related work. The problem of cold start is explained in section III. Section IV describes the strategy that will be used to mitigate the cold start latency. The Experimental setup along with the results and analysis will be discussed in Section V. We conclude the paper in Section VI.
\section{Related Work}
Serverless is a state-of-the-art field cloud computing paradigm, which has gotten tremendous attention in the last few years from the industry as well as academia. However, due to its short journey, not much literature is available that tackles the problem like cold-start.
Even though organizations rapidly migrate their applications to the cloud, the cold start problem persists. McGrath  \cite{mcgrath} and Wang \cite{wang} measured cold start latency incurred on various serverless platforms and compared them. The requests were sent in an increasing interval, from one minute to thirty minutes. It was shown that if the interval between two function calls were more than 5 minutes, it resulted in an apparent spike in the latency. Also, it was shown that cold start latency differed from one platform to another. After conducting more rigorous experiments on different platforms, Wang et al. \cite{wang} showed that the median warm start latency in Google is more than three times that in AWS, and Azure is almost thirteen times that of AWS, and the median cold start latency of AWS is more than ten times of its warm start, in Azure, it is 11.37 times. It is evident that the time required to handle a request is more for the cold start than the warm start. Some of the authors came up with innovative approaches to tackle the problem.

Some of the authors came up with innovative approaches to tackle the problem. Slacker \cite{slacker} identifies critical packages while launching a container. By prioritizing these packages and lazily loading others, it can reduce the container startup latency.
Gias and Casale \cite{gias} found out that the problem of cold start is very similar to the enhancement in the cache hit ratio. By applying the same method as in the cache hit ratio to the Faas architecture, it was seen that most of the resources were not being used. They proposed COCOA, a queuing-based approach to evaluate the effect of cold start on response times. Depending on the state of the function, variable values of memory function were taken into account. They developed a simulator, which showed that COCOA could cut down on overprovisioning by almost seventy per cent in some workloads while following the SLAs.



McGrath et al. \cite{mcgrath} proposed a scheme where there would be workers in warm and cold queues wherein old container scans are reutilized, and new ones can be created. The drawback here is that each function is mapped to only one single container. Therefore, it is not an easy task to reduce the latency caused by cold start and improve the performance of the cloud computing platform.
Ghosh et al. \cite{addya} looked at various serverless architectures and evaluated them using Lambda functions in AWS and compared the performance and the latency incurred in it with a traditional virtual machine (VM) based approach. It was seen that the time taken to access and query the database in serverless is nearly fourteen times that setup based on VM. With the help of some caching techniques, they reduced the response latency, which considerably improved the overall response time.
Daw et al. \cite{daw} tried to remove the problem of cascading cold starts, in which they introduced a novel approach Xanadu, where the resources are provisioned to the serverless platforms by speculating it beforehand. Even in standalone correlated functions, Xanadu minimizes overhead latencies when used with the implicit chain detection technique. When compared to Knative and OpenWhisk, Xanadu offers considerable performance gains, restricting cascading cold starts to a single event and minimizing the cost overruns that come with resource pre-deployments.

\section{Problem Description}
In Serverless Computing, before the actual execution, functions are containerized. This containerization process is done in two ways; first, source codes of the function are loaded onto the readily available container, and second is if readily available containers are not available regular containerization procedure is followed, which includes 
container creation, dependency loading, environment setup, network creation, and source code loading.

To prepare or store containers, computing resources are required. In serverless, if there is no incoming request for a specific time, readily available containers get deleted, and resources that handle those containers get deallocated, also known as resource scale down. Suppose a certain number function gets invoked in this situation due to the unavailability of required computing resources. In that case, the containerization process will take much more time, which will cause cold start. In another case, if the number of function invocations is more than the resources can handle, it will also cause a higher time to prepare the containers, causing cold start. 
Let us assume the set of upcoming functions, $F$, and the average execution time of $i^{th}$ function is $t_{i}$ and then the function with the added latency $T_{i}$ of cold start will as per \ref{eqn1}.
 \begin{equation}\label{eqn1}
     T_i^{Cold}=t_{i}+T_{i}
 \end{equation}
Through this paper we aim to minimize the cold start time $T_i^{Cold}$ as much as possible using learning based statistical methods.

\section{Predicting the Future Load}
In order to mitigate the cold start time, the proposed model predicts the probable future incoming function requests; this enables the model to determine how much computing resources are needed to be provisioned. The model will predict the number of upcoming future requests by using the statistical time-series model named Seasonal Autoregressive Integrated Moving Average, also known as \textit{SARIMA} \cite{farhath2016survey}. This statistical model works better when the prediction period is on a day-to-day basis and reactive to instantaneous changes, making this model a better selection over deep learning models, which are also computationally intensive concerning the used technique.
\subsection{SARIMA Model}
A SARIMA model or a Seasonal ARIMA model is given by seven whole parameters, the first three are \emph{p, d and q}. \emph{p} corresponds to the order of the AR piece, the \emph{d} to the order
of the integrated piece or how many times we take a difference to make it stationary and \emph{q} to the order of the MA piece. The next three are kind of just analogues of the first three except in a seasonal context. \emph{m} is the seasonal factor, which is the number of periods within a year it takes for the seasonality to repeat. 

After combining all these parameters, the notation for SARIMA can be represented as:

\textbf{\emph{SARIMA(data,order=(p,d,q), seasonal\_order=(P,D,Q,m))}}

In order to understand the (SARIMA) model \cite{sarima}, lets first break ARIMA into three parts. \emph{AR (denoted by $p$)} is autoregressive, which means we want to evaluate the value of our time series today, based on the value of the time series some periods in the past. \emph{I (denoted by $q$)} stands for integrated which means the time series might have some sort of upwards or downwards trend so we use differencing to make it stationary. \emph{MA (denoted by $d$)} or Moving Average means that we are taking errors from a previous time period and apply that information to the next prediction. At last, the \emph{S} stands for seasonality, which means there might be a repeating pattern in a day, week, month, year etc that happens over and over and over again over time.
For example,
\textbf{\emph{SARIMA(3,0,0)(3,0,2)24}}

The parameters \emph{P, D, and Q} change with respect to seasonality
\emph{m}. For example, an $m$ of 24 for hourly data suggests a seasonal cycle on a per day basis. A \emph{P=3}, would use the last three seasonally offset observations in the model \emph{t-(m * 1), t-(m * 2), t-(m*3)}. A \emph{D} of 0 would not take any seasonal differencing into considerations and a \emph{Q=2} would use errors of the second order in the model (e.g. moving average).

The parameters can be picked by analysing  Auto Correlation Function (ACF) and Partial Auto Correlation Function (PACF) plots by examining some of the recent time series values. These plots can also be used to get the values of seasonal components (P, D, Q) by looking at correlation at seasonal lag time steps. 

\section{Experimental Results and Analysis}
In this section we have verify the proposed method through simulated experiment. 
\subsection{Experimental Setup}
\subsubsection{System Configuration}
In order to evaluate the model and test the PBA, we create an experiment setup on top of \texttt{Docker} and \texttt{Kubernetes} on a physical node. The system configuration can be found in Table \ref{table:sysconfig}.

\begin{table}[h]
\label{table:sysconfig}
\caption{\label{table:sysconfig}System configuration}
\centering
\begin{tabular}{ |p{3cm}||p{4cm}| }
 \hline
 Processor & Intel(R) Core(TM) i5-7200U CPU @ 2.50GHz \\[1ex]
 \hline
 Memory & 8018MB\\[1ex]
 \hline
 Operating System &	Ubuntu 20.04.2 LTS \\[1ex]
 \hline
 Kernel	& Linux 5.4.0-73-generic (x86\_64)\\[1ex]
 \hline
 Kubernetes & version=v1.18.4\\[1ex]
 \hline
 Docker & version=20.10.3\\[1ex]
 \hline
\end{tabular}
\vspace{1em}
\end{table}

\begin{figure}[h]
\centerline{\includegraphics[width=0.9\linewidth]{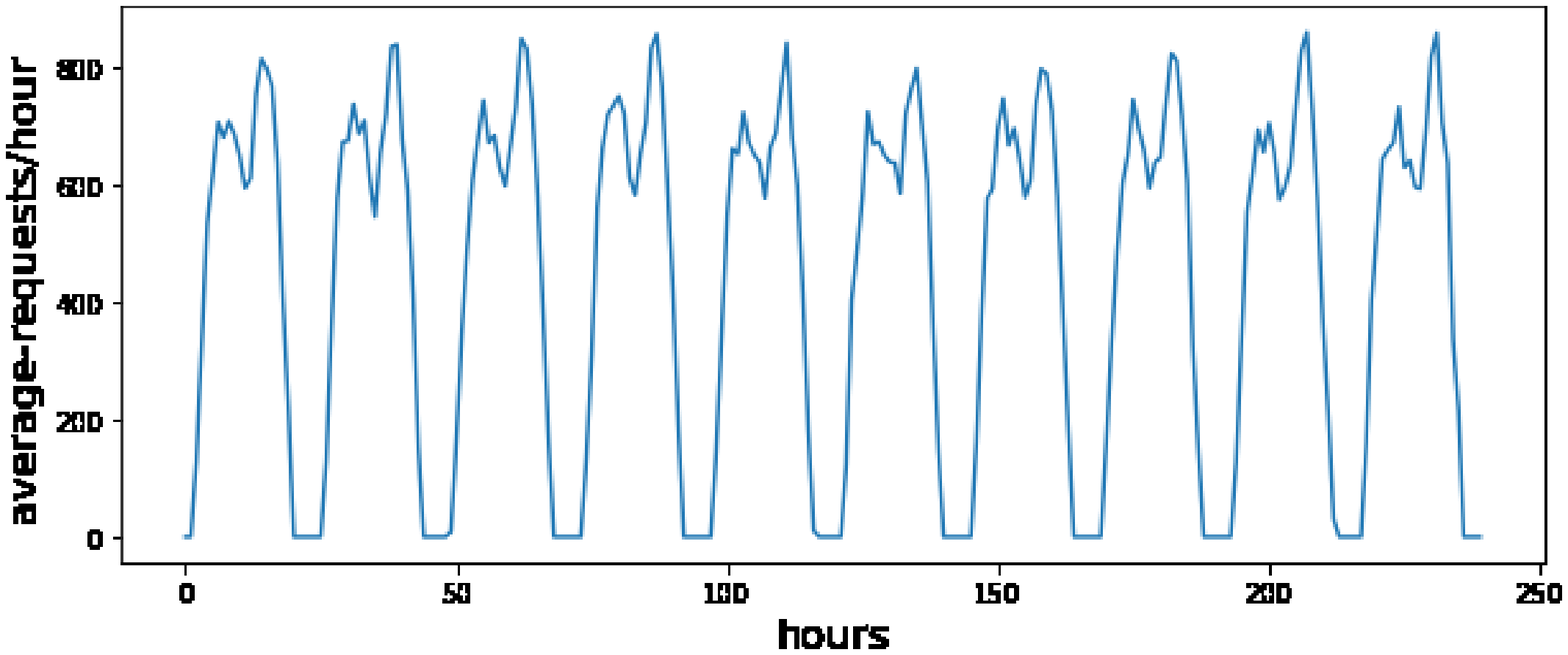}}
\caption{Dataset}
\label{fig:dataset}
\end{figure}
\subsubsection{Dataset}
Instead of using a real world data set, we have used a data set generator, which produces random data for the average number of request in an hour for any number of days. In our experiment we generated a data set for 10 days, each day containing 24 values, which denotes the average number of requests per hour for 24 hours. Figure \ref{fig:dataset} shows the data set taken into consideration.

\subsubsection{Horizontal Pod Autoscaler (HPA)}
\begin{figure}[htb]
\centering
\includegraphics[width=0.4\textwidth]{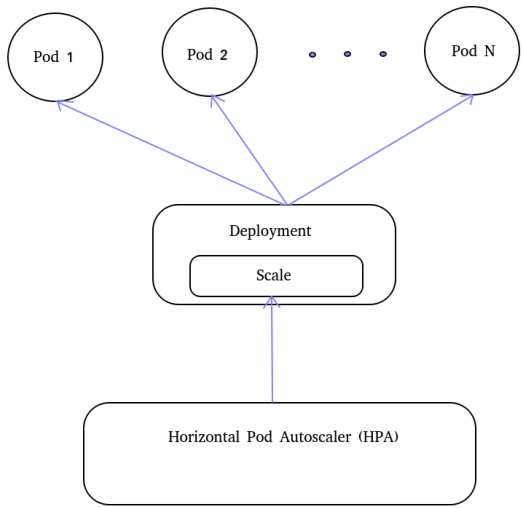}
\caption{HPA architecture.}
\label{fig:hpa}
\end{figure}

Kubernetes provides a default autoscaler which is called the Horizontal Pod Autosclaer (HPA), which scales up or scale down the pods with respect to the CPU and memory usage. This default autoscaler can be configured to an extent such that it can handle the load reasonably well. By tweaking and testing some of the parameters, HPA can give really good results and can be applied on most of the normal application. But if the application needs to be highly responsive or gets huge amount of requests in a very short period of time, then the HPA might not be able to handle it very well. There is also the case of idle time, where all the functions and containers are deleted due to no incoming requests, and are needed to be deployed again. The default HPA will almost always incur an added latency due to the default behaviour of cloud platforms of scaling the application to zero when not in use. Figure \ref{fig:hpa} shows the HPA architecture.

\subsubsection{Prediction Based Autoscaler (PBA)}
Instead of using the deciding metric for scaling as CPU or memory usage, we rely only on the prediction made by the SARIMA model and scale up or scale down in accordance with the predicted values. With the help of Kubernetes API, we propose a custom autoscaler known as the Prediction Based Autoscaler (PBA). PBA is a simple python script fed with the predicted values obtained from the SARIMA model. After some experiments, it was evident that a pod can only handle a certain number of requests for a particular set of configurations that the pod is allowed to have. Also, the amount of time required to create one more replica took around 30s. Therefore, if we knew the number of incoming requests beforehand, then with the help of some simple arithmetic, we could easily set up the required number of containers just in time to serve the incoming requests; this will potentially prevent the cold start since we have set up the environment and containers and done all the pre-processing required to serve the incoming requests. 



\begin{algorithm}
\caption{Prediction Based Autoscaler \textbf{(PBA)}}
\SetAlgoLined
 \textbf{Initialize: }first = True,  reqPerPod =int(sys.argv[1]), \\ initialDelay = int(sys.argv[2]),        interval = int(sys.argv[3]) \\
 \For{prediction in pred:}{
  \eIf{condition}{
  ct = math.ceil(prediction/reqPerPod)\;
  ft = math.ceil(prediction/reqPerPod)\;
  }{
  ft = math.ceil(prediction/reqPerPod)\;
  }
  \textbf{try}\\
        {
        \hspace{1em}scale up the number of pods with max(ft, ct)\; \\
        \hspace{1em}ct = math.ceil(prediction/reqPerPod)\; \\
        }
  \textbf{except}\\{
        \hspace{1em}print(''error")\; \\
        \hspace{1em}time.sleep(interval - (initialDelay if first else 0))\; \\
        \hspace{1em}first = False\; \\
 }
 }
\end{algorithm}
\subsubsection{Application Deployment and Monitoring}
For the purpose of testing our strategy, we created a simple \textit{nodejs} application with 3 endpoints, one static end point and two other endpoints, out of which one writes to the database and the other reads from the database. We have used \texttt{redis} as our database so that there is minimal interference due to the network latency, and we can see the actual response time of the requests that are being served. We create \textit{.yaml} files for both \textit{nodejs} and \textit{redis-db} and deploy it on the \textit{Kubernetes cluster} with the help of \textit{kubectl}, which converts the configuration written in the \textit{.yaml} files to JASON format and makes the API request to the Kubernetes. All of the deployments pertaining to the application were in the default namespace. 

In the monitoring namespace, all the components required for the purpose of monitoring the metrics generated from the application and the cluster itself are deployed. The two main monitoring deployments were \texttt{Prometheus}\footnote{https://prometheus.io/} and \texttt{Grafana}\footnote{https://grafana.com/}. \textit{Prometheus} maintains a time series database that stores all the metrics data like current CPU usage, number of incoming requests, response latency etc. \textit{Grafana} fetches the values from the time series database produced by \textit{prometheus} and makes it easy to visualize it through very cool looking graphs, gauges, charts etc in a web browser. All of the manifest can be found in the \cite{monitoring}.

\subsubsection{k6 - load generator}
\texttt{k6}\footnote{https://k6.io/} is is an open source load testing tool used to measure the performance of an application under varied amount of loads. The two common use cases of \textit{k6} is load testing and performance monitoring. There are various kinds of load testing that can be done upon a website or microservice like spike, stress and soak test. We could monitor the performance of our production environment when put under various test. The \textit{k6} load generator was also deployed on the default namespace in the \textit{Kubernetes} cluster along with the node application and \textit{redis-db}.

\subsection{Experimental Results}
\subsubsection{Actual $vs.$ Predicted Values}
\begin{figure}[htb]
\centering
\includegraphics[width=0.9\linewidth]{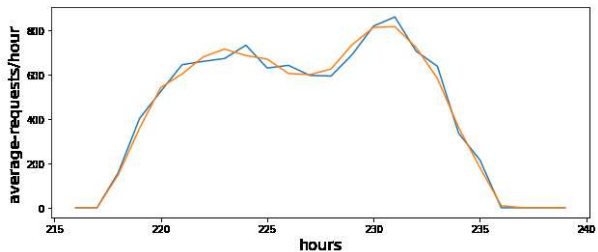}
\caption{Predicted $vs.$ Actual data.}
\label{fig:predvsactual}
\end{figure}

In order to predict the future values, we used the SARIMA model. Figure \ref{fig:predvsactual} shows the comparison between the actual and the predicted values. It can be seen that the prediction is quite accurate and close to the actual values.

\begin{figure}[ht]
\centerline{\includegraphics[width=0.9\linewidth]{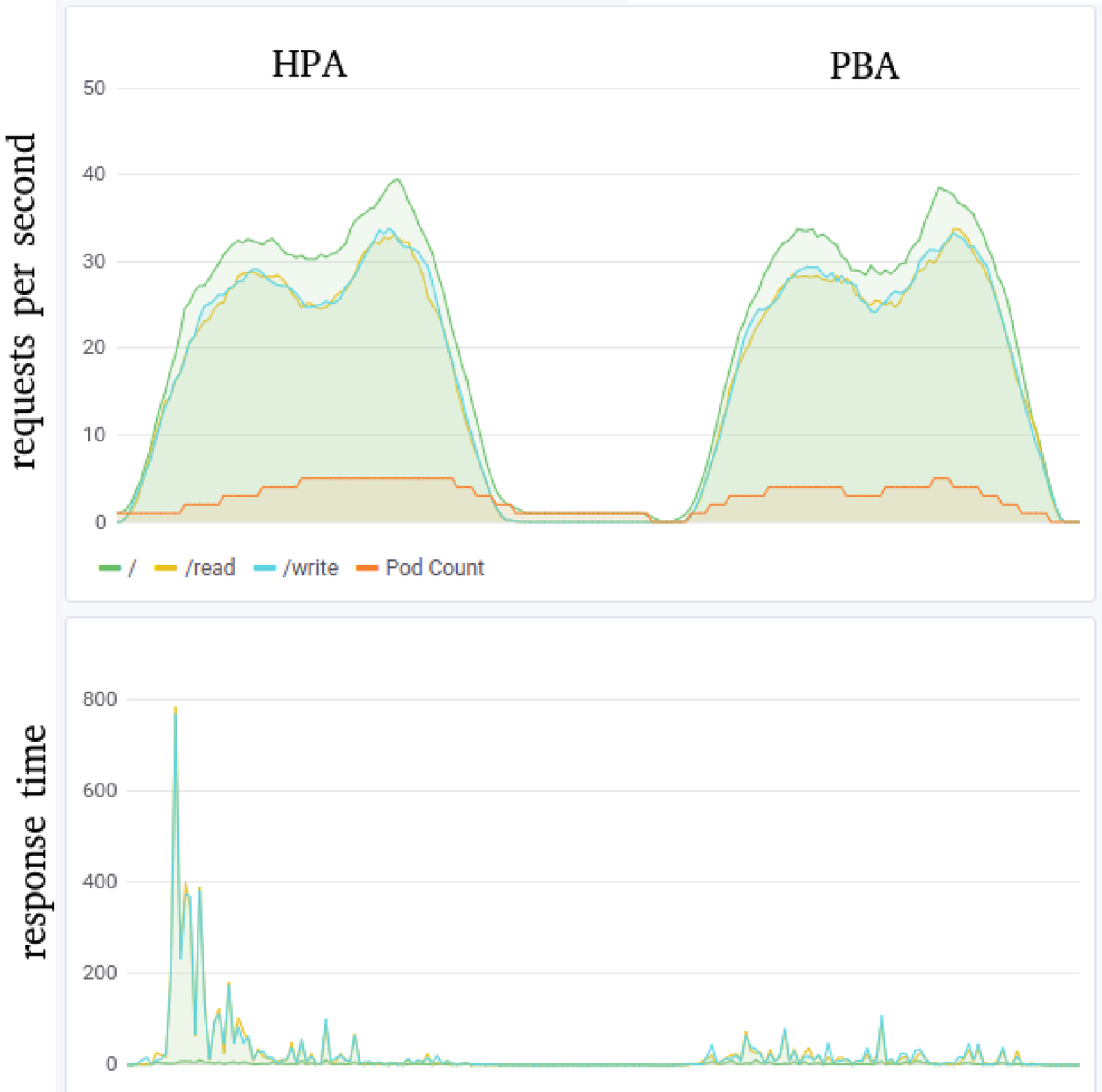}}
\caption{Comparison of the response time between HPA and PBA}
\label{fig:difference}
\end{figure}
\subsubsection{Average Response Time per Second}
In order to evaluate our strategy of using a classic statistical model like SARIMA for predicting the incoming requests along with PBA, we compare it with the default autoscaler of \textit{Kubernetes}, HPA. Firstly, we show the average response time taken to execute a request. The number of requests sent to the node are proportional to the number of requests an Indian payments service company gets during the 24 hours of the day. The results can be seen in Figure \ref{fig:difference}, depicting a \textit{Grafana} dashboard that compares the \textit{requests/sec} along with the average \textit{response time/sec} of both PBA and HPA.

\subsubsection{Pod Count Comparison}
\begin{figure}[ht]
\centerline{\includegraphics[width=\linewidth]{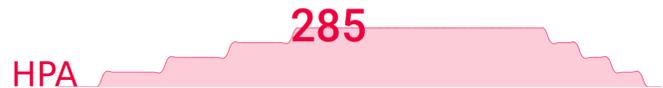}}
\caption{HPA pod count}
\label{fig:hpapodcount}
\end{figure}

\begin{figure}[htb]
\centerline{\includegraphics[width=\linewidth]{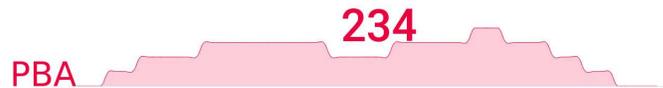}}
\caption{PBA pod count}
\label{fig:pbapodcount}
\end{figure}

    Upon conducting the experiment numerous times, it was found out that PBA performed better than HPA in terms of the number of pods being spawned. Figure. \ref{fig:hpapodcount} and Figure. \ref{fig:pbapodcount} shows the number of pods that are being consumed by HPA and PBA respectively. Since the amount of resources used is proportional to the pod count, we can infer that PBA 18\% less resources that HPA.

\begin{figure}[htb]
\centerline{\includegraphics[width=\linewidth]{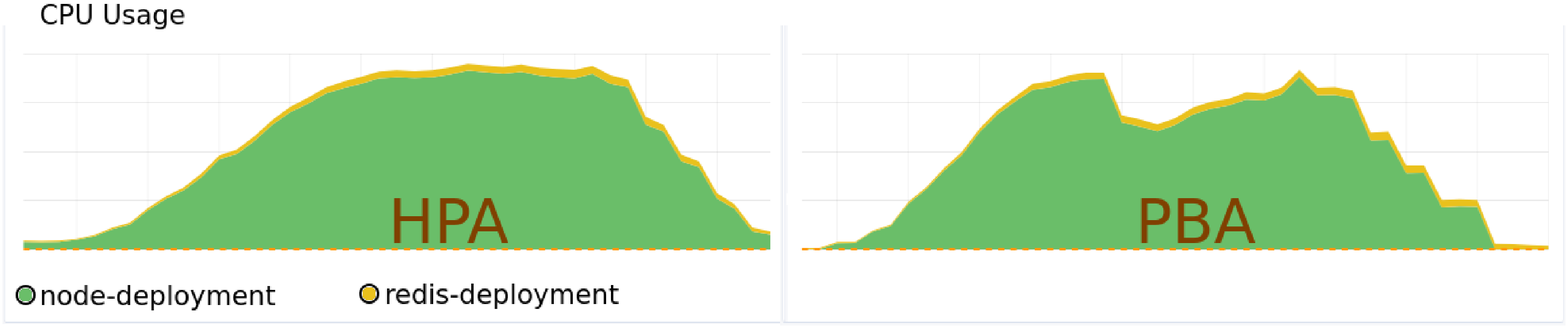}}
\caption{Comparison of the response time between HPA and PBA}
\label{fig:hpavspbacpu}
\end{figure}
\begin{figure}[htb]
\centerline{\includegraphics[width=\linewidth]{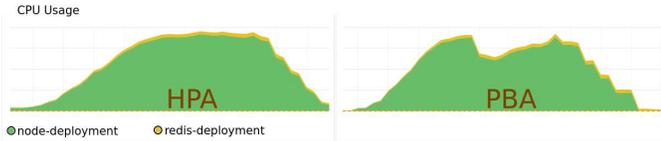}}
\caption{Comparison of the response time between HPA and PBA}
\label{fig:hpavspbamemory}
\end{figure}

\subsubsection{Overall CPU and Memory consumption}
Figure. \ref{fig:hpavspbacpu} and Figure. \ref{fig:hpavspbamemory} compare the CPU and memory usage respectively for both \textit{nodejs} and \textit{redis-db} deployment. HPA scales up or scales down according to the percentage of CPU used, whereas PBA has the ability to scale up or scale down the number of pods on the basis of the prediction made earlier with the help of SARIMA model. The same logic is applicable for the amount of memory used. The consumption of CPU and memory is found to be much lower in PBA than in HPA.

\section{Conclusion and Future Work}
    Serverless computing is being adopted by a lot of applications, but there is still a major chunk who are not able to use the benefits of serverless computing due to the problem of cold start. It is very well proven that a cold start takes significant amount of time more, than invoking a function whose container is already deployed. The challenging part is to reduce the cold start latency without the consumption of extra resources.
    In this paper, we intended to reduce the cold start latency as much as possible, but not at the expense of additional resources. We propose to predict the future incoming requests with the help of a classical statistical model called SARIMA. After getting the predictions from the model, we feed it into the PBA, which scales up or scales down the number of pods in accordance with the predictions made by the model. This facilitates the provisioning of roughly exact number of resources that are required by the application at any given point in time. Therefore, there is no over-provisioning or under-provisioning of resources, and the application runs in a smooth manner. In order to validate our strategy we compare our PBA with HPA, which is the default pod autoscaler in Kubernetes. Upon comparing the two, HPA suffered from cold start whereas PBA was able to subdue the effect of cold start in a graceful manner. PBA also consumed 18\% less resources than HPA.
    
    There still remain many aspects of cold start in serverless computing that need to be better understood and refined. Future predictions is one of the ways to tackle this problem, but a significant amount of research is required to better understand the core of the problem and if there are any innovative solutions to it.
    Though cold starts have been in the lime light lately in the world of serverless, they are little known outside the spheres
    of cloud computing. Nevertheless, research in this field is going on strong, and hopefully, In the coming years, the problem of cold start would perish.

\bibliographystyle{IEEEtran}
\bibliography{ref}

\begin{thebibliography}{10}
\providecommand{\url}[1]{#1}
\csname url@samestyle\endcsname
\providecommand{\newblock}{\relax}
\providecommand{\bibinfo}[2]{#2}
\providecommand{\BIBentrySTDinterwordspacing}{\spaceskip=0pt\relax}
\providecommand{\BIBentryALTinterwordstretchfactor}{4}
\providecommand{\BIBentryALTinterwordspacing}{\spaceskip=\fontdimen2\font plus
\BIBentryALTinterwordstretchfactor\fontdimen3\font minus
  \fontdimen4\font\relax}
\providecommand{\BIBforeignlanguage}[2]{{%
\expandafter\ifx\csname l@#1\endcsname\relax
\typeout{** WARNING: IEEEtran.bst: No hyphenation pattern has been}%
\typeout{** loaded for the language `#1'. Using the pattern for}%
\typeout{** the default language instead.}%
\else
\language=\csname l@#1\endcsname
\fi
#2}}
\providecommand{\BIBdecl}{\relax}
\BIBdecl

\bibitem{baldini2017serverless}
I.~Baldini, P.~Castro, K.~Chang, P.~Cheng, S.~Fink, V.~Ishakian, N.~Mitchell,
  V.~Muthusamy, R.~Rabbah, A.~Slominski \emph{et~al.}, ``Serverless computing:
  Current trends and open problems,'' in \emph{Research advances in cloud
  computing}.\hskip 1em plus 0.5em minus 0.4em\relax Springer, 2017, pp. 1--20.

\bibitem{addya2021comcloud}
S.~K. Addya, A.~Satpathy, B.~C. Ghosh, S.~Chakraborty, S.~K. Ghosh, and S.~K.
  Das, ``{C}o{MCLOUD}: Virtual machine coalition for multi-tier applications
  over multi-cloud environments,'' \emph{IEEE Transactions on Cloud Computing},
  to appear, 2022, {D}OI:10.1109/TCC.2021.3122445.

\bibitem{kim2020automated}
Y.~K. Kim, M.~R. HoseinyFarahabady, Y.~C. Lee, and A.~Y. Zomaya, ``Automated
  fine-grained cpu cap control in serverless computing platform,'' \emph{IEEE
  Transactions on Parallel and Distributed Systems}, vol.~31, no.~10, pp.
  2289--2301, 2020.

\bibitem{ishakian2018serving}
V.~Ishakian, V.~Muthusamy, and A.~Slominski, ``Serving deep learning models in
  a serverless platform,'' in \emph{2018 IEEE International Conference on Cloud
  Engineering (IC2E)}.\hskip 1em plus 0.5em minus 0.4em\relax IEEE, 2018, pp.
  257--262.

\bibitem{benedict2020serverless}
S.~Benedict, ``Serverless blockchain-enabled architecture for iot societal
  applications,'' \emph{IEEE Transactions on Computational Social Systems},
  vol.~7, no.~5, pp. 1146--1158, 2020.

\bibitem{wang2020supporting}
I.~Wang, E.~Liri, and K.~Ramakrishnan, ``Supporting iot applications with
  serverless edge clouds,'' in \emph{2020 IEEE 9th International Conference on
  Cloud Networking (CloudNet)}.\hskip 1em plus 0.5em minus 0.4em\relax IEEE,
  2020, pp. 1--4.

\bibitem{rahman2019serverless}
M.~M. Rahman and M.~H. Hasan, ``Serverless architecture for big data
  analytics,'' in \emph{2019 Global Conference for Advancement in Technology
  (GCAT)}.\hskip 1em plus 0.5em minus 0.4em\relax IEEE, 2019, pp. 1--5.

\bibitem{baresi2019towards}
L.~Baresi and D.~F. Mendon{\c{c}}a, ``Towards a serverless platform for edge
  computing,'' in \emph{2019 IEEE International Conference on Fog Computing
  (ICFC)}.\hskip 1em plus 0.5em minus 0.4em\relax IEEE, 2019, pp. 1--10.

\bibitem{wang2019distributed}
H.~Wang, D.~Niu, and B.~Li, ``Distributed machine learning with a serverless
  architecture,'' in \emph{IEEE INFOCOM 2019-IEEE Conference on Computer
  Communications}.\hskip 1em plus 0.5em minus 0.4em\relax IEEE, 2019, pp.
  1288--1296.

\bibitem{lin2020modeling}
C.~Lin and H.~Khazaei, ``Modeling and optimization of performance and cost of
  serverless applications,'' \emph{IEEE Transactions on Parallel and
  Distributed Systems}, vol.~32, no.~3, pp. 615--632, 2020.

\bibitem{vahidinia2020cold}
P.~Vahidinia, B.~Farahani, and F.~S. Aliee, ``Cold start in serverless
  computing: Current trends and mitigation strategies,'' in \emph{2020
  International Conference on Omni-layer Intelligent Systems (COINS)}.\hskip
  1em plus 0.5em minus 0.4em\relax IEEE, 2020, pp. 1--7.

\bibitem{manner2018cold}
J.~Manner, M.~Endre{\ss}, T.~Heckel, and G.~Wirtz, ``Cold start influencing
  factors in function as a service,'' in \emph{2018 IEEE/ACM International
  Conference on Utility and Cloud Computing Companion (UCC Companion)}.\hskip
  1em plus 0.5em minus 0.4em\relax IEEE, 2018, pp. 181--188.

\bibitem{Kubernetes}
Kubernetes. (2022) Https://kubernetes.io/docs/home/.

\bibitem{mcgrath}
G.~McGrath and P.~R. Brenner, ``Serverless computing: Design, implementation,
  and performance,'' in \emph{2017 IEEE 37th International Conference on
  Distributed Computing Systems Workshops (ICDCSW)}, 2017, pp. 405--410.

\bibitem{wang}
L.~Wang, M.~Li, Y.~Zhang, T.~Ristenpart, and M.~Swift, ``Peeking behind the
  curtains of serverless platforms,'' in \emph{2018 {USENIX} Annual Technical
  Conference ({USENIX} {ATC} 18)}.\hskip 1em plus 0.5em minus 0.4em\relax
  Boston, MA: {USENIX} Association, Jul. 2018, pp. 133--146.

\bibitem{slacker}
T.~Harter, B.~Salmon, R.~Liu, A.~C. Arpaci-Dusseau, and R.~H. Arpaci-Dusseau,
  ``Slacker: Fast distribution with lazy docker containers,'' in \emph{14th
  {USENIX} Conference on File and Storage Technologies ({FAST} 16)}.\hskip 1em
  plus 0.5em minus 0.4em\relax Santa Clara, CA: {USENIX} Association, Feb.
  2016, pp. 181--195.

\bibitem{gias}
A.~U. Gias and G.~Casale, ``Cocoa: Cold start aware capacity planning for
  function-as-a-service platforms,'' in \emph{2020 28th International Symposium
  on Modeling, Analysis, and Simulation of Computer and Telecommunication
  Systems (MASCOTS)}, 2020, pp. 1--8.

\bibitem{addya}
B.~C. Ghosh, S.~K. Addya, N.~B. Somy, S.~B. Nath, S.~Chakraborty, and S.~K.
  Ghosh, ``Caching techniques to improve latency in serverless architectures,''
  in \emph{2020 International Conference on COMmunication Systems NETworkS
  (COMSNETS)}, 2020, pp. 666--669.

\bibitem{daw}
N.~Daw, U.~Bellur, and P.~Kulkarni, ``Xanadu: Mitigating cascading cold starts
  in serverless function chain deployments,'' in \emph{Proceedings of the 21st
  International Middleware Conference}, ser. Middleware '20.\hskip 1em plus
  0.5em minus 0.4em\relax New York, NY, USA: Association for Computing
  Machinery, 2020, p. 356–370.

\bibitem{farhath2016survey}
Z.~A. Farhath, B.~Arputhamary, and L.~Arockiam, ``A survey on arima forecasting
  using time series model,'' \emph{Int. J. Comput. Sci. Mobile Comput}, vol.~5,
  pp. 104--109, 2016.

\bibitem{sarima}
{Seasonal Autoregressive Integrated Moving Average}. (2021) Sarima. Available:
  \url{https://machinelearningmastery.com/time-series-forecasting-methods-in-python-cheat-sheet/}.

\bibitem{monitoring}
{Github repository}. (2021) monitoring. Available:
  \url{https://github.com/prometheus-operator/kube-prometheus/tree/main/manifests}.

\end{thebibliography}

\end{document}